\documentclass{elsart1p}
\usepackage{epsfig}
\begin{document}
\runauthor{Marco Battaglia}
\begin{frontmatter}
\title{Tracking and Vertexing\\ with a Thin CMOS Pixel Beam Telescope}
\author[UCB,LBNL]{Marco Battaglia},
\author[INFN]{Dario Bisello},
\author[PURDUE]{Gino Bolla},
\author[PURDUE]{Daniela Bortoletto},
\author[LBNL]{Devis Contarato},
\author[LBNL,INFNPV]{Silvia Franchino},
\author[LBNL,INFN]{Piero Giubilato},
\author[UCB,LBNL]{Lindsay Glesener},
\author[UCB,LBNL]{Benjamin Hooberman}
\author[INFN]{Devis Pantano},
\address[UCB]{Department of Physics, University of California at 
Berkeley, CA 94720, USA}
\address[LBNL]{Lawrence Berkeley National Laboratory, 
Berkeley, CA 94720, USA}
\address[INFN]{Dipartimento di Fisica, Universita' di Padova and INFN,
Sezione di Padova, I-35131 Padova, Italy}
\address[PURDUE]{Purdue University, Department of Physics, West Lafayette, 
IN 47907, USA}
\address[INFNPV]{Dipartimento di Fisica, Universita' di Pavia and INFN,
Sezione di Pavia, I-35131 Pavia, Italy}
\begin{abstract}
We present results of a study of charged particle track and vertex 
reconstruction with a beam telescope made of four layers of 50~$\mu$m-thin 
CMOS monolithic pixel sensors using the 120~GeV protons at the FNAL 
Meson Test Beam Facility. We compare our results to the performance
requirements of a future $e^+e^-$ linear collider in terms of particle 
track extrapolation and vertex reconstruction accuracies.
\end{abstract}
\end{frontmatter}

\typeout{SET RUN AUTHOR to \@runauthor}


\section{Introduction}

The physics program at a high energy linear $e^+e^-$ collider, such as 
the ILC~\cite{Djouadi:2007ik} or CLIC~\cite{Battaglia:2004mw}, relies 
significantly on the jet flavour tagging capability of its detectors. 
Distinguishing $b$, $c$ and light quark jets as well as 
$\tau$ leptons is of prime importance both for extracting signals of new 
particles, which may couple preferentially to heavy fermions, and for analysing 
new sectors of particle physics, such as the Higgs mechanism. The identification 
of heavy quarks and leptons largely relies on their long lifetimes, which result 
in secondary and tertiary vertices separated from the beam interaction point by 
distances which range from few hundred microns to several millimetres, or even 
centimetres at multi-TeV collision energies. In order to efficiently identify 
these secondary vertices, it is essential to achieve excellent resolution on the 
extrapolation of charged particle tracks to their production point, 
$\sigma_{{\mathrm{extr}}}$. 
The requirement for a 0.5~TeV - 1.0~TeV linear collider has been identified 
to be  $\sigma_{{\mathrm{extr}}} = 5 \oplus \frac{10}{p_t}$~($\mu$m), where $p_t$ 
is the momentum component in the transverse plane, measured in GeV. These 
asymptotic and multiple scattering terms are respectively 3 and 8 times better 
than the design performance of the LHC detectors and 1.5 and 3 times better 
than the best extrapolation resolution ever achieved by a collider experiment 
detector, that of the VXD3 vertex detector in SLD at the Stanford Linear 
Collider~\cite{vxd3}. Achieving these performances requires the development of 
novel sensor technologies. Monolithic silicon pixellated sensors, back-thinned 
to 20-50~$\mu$m 
have emerged as possibly the most appealing solution and a significant R\&D 
effort is addressing various technologies and architectures in the framework 
of the world-wide linear collider studies. Small pixel size, fast readout and 
limited power dissipation, to enable airflow cooling and remove the material 
burden of active cooling, are some of the main issues being addressed. This 
R\&D is guided by the physics requirements and detailed simulation studies 
of tracking performances. It is essential to validate these simulations with 
data collected from prototype trackers made of thin pixel sensors with a geometry 
resembling that envisaged for the ILC, under realistic operating conditions and 
hit occupancy levels. 

In this paper we report the first results of the T966 beam test experiment 
at FNAL, which aims at a detailed study of tracking and vertexing capability 
of a prototype tracker based on thin CMOS monolithic Si pixel sensors.

\section{The T966 Thin Pixel Telescope}

The T966 beam test experiment took data at the Meson Test Beam Facility (MTBF) 
at FNAL in the Summer 2007 and deployed a beam telescope made of four layers
of thinned CMOS pixel sensors. A prototype thin pixel  telescope (TPPT-1) had 
been successfully operated on the 1.5~GeV BTS electron beam-line at the LBNL 
Advanced Light Source~\cite{Battaglia:2006tf}.
The T966 beam telescope has better mechanical positioning of the detector 
layers compared to the TPPT-1 condiguration and also includes remotely controlled 
stages for positioning a detector under test 20~mm downstream from the telescope 
and inserting a copper target $\simeq$~30~mm upstream. The telescope consists of 
four detector layers, spaced 15~mm apart. This provides a geometry quite similar 
to that envisaged for the ILC Vertex Tracker, which has five to six sensitive layers 
spaced by $\simeq$~11~mm, located at radii from 15~mm to 60~mm~\cite{ilcvtx}. 
In particular, by extrapolating tracks reconstructed from hits on layers 2, 3 and 4 
onto layer 1, we can study the extrapolation resolution for particle tracks over 
the same lever arm foreseen at the ILC, where the first layer of the Vertex Tracker 
would be positioned $\simeq$~15~mm away from the beam interaction point. 

The telescope uses the MIMOSA-5 chip, developed by IPHC, Strasbourg 
(France)~\cite{mimosa, deptuch-ref}.
This chip, fabricated in the AMS 0.6~$\mu$m process, features a large
active area of 1.7$\times$1.7~cm$^2$ and more than 1~Million pixels, 
arranged in four independent sectors. The epitaxial layer is 14~$\mu$m 
thick and the pixel size 17$\times$17~$\mu$m$^2$.
Detectors have been thinned by Aptek Industries~\cite{aptek-ref} to 
(50$\pm$7)~$\mu$m using a proprietary grinding process.
Each detector is glued to a PC board using a precision mounting jig which 
ensures a position accuracy of the detector corners w.r.t.\ the boards of 
$\simeq$~20~$\mu$m. The board has a cut-out area corresponding to the sensor 
sensitive region to reduce multiple scattering. We verify the flatness of the 
thin chips after gluing them on the PC board using an optical survey system which
has a resolution of $\simeq$~1~$\mu$m. We measure the deviation between the height 
of 100~points on the chip surface and that of a plane interpolated through them. 
This has a Gaussian width of 7~$\mu$m.
The PC boards are mounted on the readout board and held in place by precision 
mechanics which ensures their relative alignment. We verify the stability by 
measuring fiducial reference marks on the four PC boards before and after 
mechanically disturbing the setup. From these measurements we estimate that the 
sensitive area of each layer is stable within $\le$~20~$\mu$m in the $z$ 
coordinate, along the beam axis, and $\le$~10~$\mu$m in the $x$ and $y$ coordinates, 
transverse to it. The telescope and its readout electronics are installed in an 
optical enclosure. The system is aligned on the beam-line using a laser beam and 
is operated at a constant temperature of +20$^{\circ}$C, by flowing cold air 
inside the optical enclosure. The temperature is monitored by a thermocouple 
installed between layer 2 and 3 of the telescope.

One sector of each chip, corresponding to a 510$\times$512 pixel array, 
is read out through a custom FPGA-driven acquisition board.
The trigger is provided by two finger scintillators, S1 and S2, mounted just 
in front of the optical enclosure, defining a fiducial area of 
$\simeq$1~cm$^2$ around the sector of the MIMOSA chip that is 
readout. The S1-S2 coincidence is gated on the 4~s-long spill extraction 
signal. The readout cycle consists of a reset followed by the readout of two 
subsequent frames of the 510$\times$512 pixel sector. The chip is read out at 
3.125~MHz, which corresponds to 84~ms for reading a full sector. Four 14~bits, 
40~MSample/s ADCs simultaneously read the four sensors, while an array of 
digital buffers drives all the required clocks and synchronisation signals. 
The FPGA has been programmed to generate the clock pattern and collect the 
sampled data from the ADCs. A 32~bit wide bus connects the FPGA to a
digital acquisition board installed on a control PC. Due to the slow readout 
and data transfer, the telescope acquires 7 to 10~events/spill.
Data are processed on-line by a LabView-based program, which performs correlated 
double sampling and pedestal subtraction, noise computation and cluster identification.  
To reduce the amount of data written to disk, only the addresses and pulse heights of
the pixels in a fixed matrix around the centre of a cluster candidate having a seed 
pixel with signal-to-noise ratio above~4 are recorded. Data are converted offline 
in the {\tt LCIO} format~\cite{Gaede:2003ip} for the subsequent analysis.

The average pixel noise has been measured to be (85 $\pm$ 16)~$e^-$ of equivalent 
noise charge (ENC), which is due both to the noise of the readout electronics and 
to leakage current from operating the detector at +20$^{\circ}$C. A test performed 
in the lab shows that the noise drops to $\simeq$~50~~$e^-$ of ENC, when operating 
the detector below +5$^{\circ}$C.

During the 2007 data taking, the 120~GeV proton beam was extracted from the Main 
Injector and attenuated through a pinhole collimator. The average beam size was 
measured to be about 4~mm and 9~mm in the horizontal and vertical coordinate, 
respectively, using two sets of multi-wire chambers located upstream from the 
telescope.
The average beam divergence, measured from the slope of tracks reconstructed in 
the telescope, is 0.52~mrad. The beam intensity was varied during data taking and 
resulted in an average hit density ranging from 0.07~hits~mm$^{-2}$ to 0.70~hits~mm$^{-2}$,
with local densities of up to 5~hits~mm$^{-2}$.
For comparison, in the core of hadronic jets from $e^+e^-$ collisions, the average hit 
density is expected to be in the range 0.2-1.0~hits~mm$^{-2}$ for $\sqrt{s}$=0.5~TeV 
and 0.5-2.5~hits~mm$^{-2}$ for 3.0~TeV~\cite{Battaglia:2003kn}.
The machine-induced pair background contributes $\simeq$0.05~~hits~mm$^{-2}$~bunch 
crossing$^{-1}$ at 15~mm radius at 0.5~TeV~\cite{Buesser:2005iu,ilcvtx} and  
$\simeq$0.005~~hits~mm$^{-2}$~bunch crossing$^{-1}$ at 30~mm radius at 
3.0~TeV~\cite{Battaglia:2004mw} for solenoidal fields of 4~T and 5~T, respectively. 
Local densities up to 5~hits~mm$^{-2}$ are quite 
comparable to those expected in a linear collider vertex tracker. At the same time 
our redundancy is lower than that provided by a five-layered vertex tracker and a
main tracker, which makes the overall occupancy conditions quite realistic.
 
\section{Reconstruction and Simulation}

The offline analysis is performed using a set of dedicated processors developed 
for the {\tt Marlin} framework~\cite{Gaede:2006pj}. Individual processors are used 
for clustering and hit reconstruction, pattern recognition and track fit and for 
vertexing. An additional processor provides signal simulation of the pixels. 
The T966 setup is modelled by a simulation program based on the {\tt Geant-4} 
package~\cite{Agostinelli:2002hh}, which generates particle points of impact and 
energy deposits in the sensitive detector planes and accounts for the material 
surrounding the beam-line and the detectors. Simulated data are stored in {\tt LCIO} 
format and used as input to the CMOS pixel simulation processor, {\tt PixelSim}, 
implemented in {\tt Marlin}~\cite{Battaglia:2007eu}. 

\begin{figure}
\begin{center}
\epsfig{file=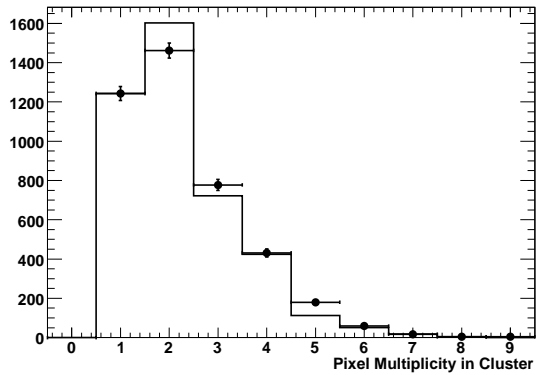,width=7.5cm}
\end{center}
\caption[]{Pixel multiplicity for clusters 
associated to a particle track. 
Data are shown as points with error bars 
and simulation with the histogram.}
\label{fig:clusters}
\end{figure}
Hits are reconstructed from the recorded pixel pulse heights as follows: 
each detector is scanned for pixels with pulse height over a given signal-to-noise 
(S/N) threshold. These are designated as cluster `seeds'. Seeds are then sorted 
according to their pulse height values and the surrounding, neighbouring pixels 
are tested for addition to the cluster. The S/N threshold is 5 for the seed pixel 
and 2.5 for the neighbour pixels.
The neighbour search is performed in a 5$\times$5 matrix.
Clusters are not allowed to overlap, i.e. pixels already associated to one
cluster are not considered for populating another cluster around a different
seed. Finally, we require that clusters are not discontinuous, i.e.\ pixels
associated to a cluster cannot be interleaved by any pixel below the neighbour 
threshold. Pixels with noise exceeding two times the average sector noise or 
giving a pulse height above the seed threshold in more than 10~\% of the events 
of a given run are flagged as noisy and removed offline for the whole run. Columns 
of pixels with more than 25 pixels having pulse height above the seed threshold in 
one event are also flagged as noisy and removed for that single event. 
Selected clusters have an average pixel multiplicity of 2.35 and average 
signal-to-noise ratio of 10.3. 
These results are in agreement with those predicted by simulation, as shown in 
Figure~\ref{fig:clusters}. Simulation predicts single point resolutions, 
$\sigma_{\mathrm{point}}$, of (2.90$\pm$0.02)~$\mu$m, (2.08$\pm$0.02)~$\mu$m and 
(1.70$\pm$0.02)~$\mu$m for cluster S/N values of 10, 15 and 20. This last value agrees 
with the resolution of 1.7~$\mu$m, reported in~\cite{mimosa}, obtained by operating 
the detector cold. The track reconstruction efficiency is estimated using simulation
where the noise of each individual layer is tuned on the data. Particle densities 
comparable to those registered in the data are simulated and pixels are masked to 
reproduce the effect of noisy pixels and columns. Simulation reproduces well the 
number of reconstructed particles and the number of associated hits. Due to the 
limited pixel S/N performance achieved in our 2007 data taking, the estimated 
track reconstruction efficiency, after applying the full set of selection criteria 
as in real data, is 0.69.

\section{Alignment}

\begin{figure}
\begin{center}
\epsfig{file=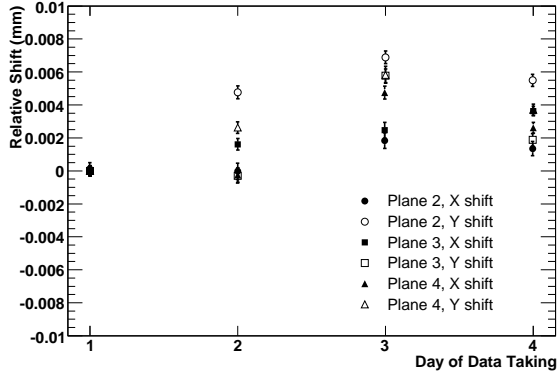,width=7.5cm}
\end{center}
\caption[]{Relative position of the telescope layers w.r.t.\ the first layer 
as a function of day of data taking as obtained from the track alignment.}
\label{fig:align}
\end{figure}

The relative position of the four layers of the telescope is obtained by using 
the result of an optical survey and a track-based alignment procedure to align the 
telescope planes with respect to a common reference system. 
The origin of this system coincides with the centre 
of the detector chip in the first layer and the $x$ and $y$ directions are aligned 
along the pixel rows and columns in their readout direction. 
For each telescope layer we consider six degrees of freedom: the three offsets along 
the $x$, $y$ and $z$ axes and the three rotations around them.
These parameters are determined from a $\chi^2$ fit which minimises the distance
between the hit positions measured on each plane and those predicted by the
extrapolation from the other planes. The optical survey provides with the 
starting positions. The track-based alignment procedure aligns the detector planes in 
pairs. We start with aligning the second telescope plane
with respect to the first. The third plane is then aligned on the extrapolated 
positions determined from the track segments fitted with the first two planes.
Finally, the fourth sensor is aligned with respect to the tracks obtained from a linear
fit between the first three telescope planes. After this initial procedure, any given
plane can be re-aligned on the extrapolations of the tracks fitted with the other three
planes. Quality cuts are applied on the track slopes and fit $\chi^2$.
Due to a $\simeq$ 5-9$^{\circ}$C difference in temperature
between the day, when the telescope is operated, and the night, when its power 
is switched off, the procedure is repeated for each day of data taking, 
in order to account for possible distortions due to thermal effects. 
Changes in the relative position of the layers of up to 6-7~$\mu$m are 
observed, as shown in Figure~\ref{fig:align}. This result is consistent with the 
measured stability of the mechanical mounting. A sub-micron accuracy on the 
relative position of any pairs of layers is obtained with samples of 
$\le$~250~fitted tracks. These results are relevant for estimating the amount
of data needed for aligning a single ladder in a full Vertex Tracker at a collider.
Alignment parameters are stored in a conditions database using the Linear 
Collider Conditions Data ({\tt LCCD}) toolkit~\cite{Gaede:2006pj}, 
a C++ based {\tt LCIO} framework. {\tt LCCD} offers the advantage of full 
database functionality while being straightforward and easy to implement.  
Conditions data consisting of a set of alignment parameters for each day 
of data taking are stored in an {\tt LCIO} file and are accessed by a 
{\tt Marlin} module at run time.  

\section{Tracking Results}

\begin{figure}
\begin{center}
\epsfig{file=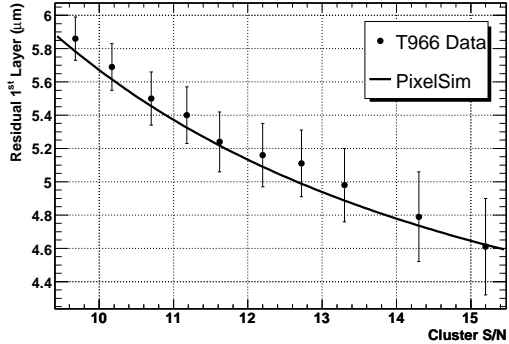,width=7.5cm,clip}
\end{center}
\caption[]{Gaussian width of the distribution of residual between track extrapolation and 
position of closest hit on the first layer of the telescope as a function of the average 
S/N of the hit clusters used for reconstructing the track. Data are shown as points with 
error bars and simulation by the continuous line.}
\label{fig:residuals_sn}
\end{figure}

Particle tracks are fitted using the reconstructed hits with a simple straight 
track model. The pattern recognition starts from hits on a given detector 
layer and looks for sets of aligned hits. Hit pairs are used as track seeds 
provided the slope does not exceed 0.0015. Additional hits are tested if their
residual w.r.t\ the track extrapolation does not exceed 35~$\mu$m. Hit bundles
corresponding to the $\chi^2$ minimum for a given hit on the starting layer are
used to define a track candidate provided that $\chi^2 <$ 25 and the slope 
does not exceed 0.0010. The slope cut removes lower momentum particles originating 
from interactions and scattering upstream in the beamline as well as badly 
reconstructed particle tracks resulting from pattern recognition failures.
The procedure is repeated using all planes. 
Ambiguities in hit association are then resolved based on the number of 
hits in a track candidate and its fit $\chi^2$. Tracks are fitted and associated 
hits are flagged. The remaining unassociated hits are used for a second pass pattern 
recognition, where three- and two-hit tracks are reconstructed from the hit 
combinations giving the minimum $\chi^2$ provided they do not exceed 25 or the 
minimum slope, respectively. The procedure is tested on simulated events with the 
same track density as the data. The fraction of correctly associated hits is 
0.93$\pm$0.01 for tracks with at least three hits. Residuals are computed from the 
distance between the track extrapolation on a test detector plane and the closest hit, 
which is not included in the track fit. Only tracks having an associated hit on the 
layer closest to the extrapolation plane are considered.
\begin{figure}
\begin{center}
\epsfig{file=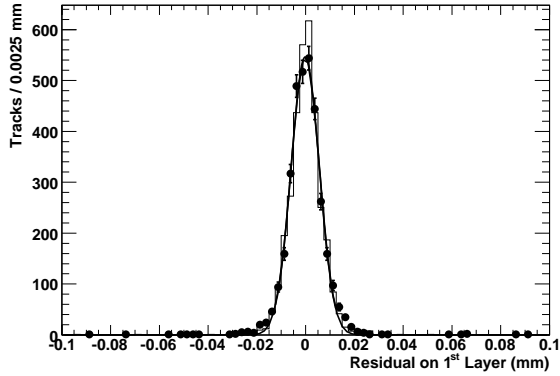,width=7.5cm,clip}
\end{center}
\caption[]{Residual between the track extrapolation and the position of the closest 
hit on the first layer of the telescope. Data are shown as points with error bars and 
simulation by the histogram. The Gaussian curve fitted to the data has a width of 
5.7~$\mu$m.}
\label{fig:residuals}
\end{figure}
First we determine the detector single point resolution by including the hits on all four 
telescope layers in the track fit. Using the fact that the layers are equally spaced, we 
can extract the single point resolution $\sigma_{\mathrm{point}}$ from the width of the 
measured residual distribution, corrected by a geometrical factor, under the assumption that 
$\sigma_{\mathrm{point}}$ is the same for all layers. We measure a Gaussian width of 
(1.57$\pm$0.02)~$\mu$m and (1.26$\pm$0.10)~$\mu$m using hits from clusters with an 
average S/N ratio of 10 and 15, respectively. Using the correction factor of 1.81 for 
our geometry, we obtain an estimate of $\sigma_{\mathrm{point}}$ = (2.85$\pm$0.04)~$\mu$m 
and (2.29$\pm$0.18)~$\mu$m for cluster hits with average signal-to-noise ratio of 10 and 
15, respectively. This result is in good agreement with the prediction of our CMOS sensor 
simulation, as discussed above.
Then we extrapolate the tracks on the second layer and study the distribution of the residuals
between the position of extrapolated and that of the closest hit, which is not included
in the track fit. This distribution has a Gaussian width of (4.6$\pm$0.2)~$\mu$m for data 
and (4.1$\pm$0.2)~$\mu$m for simulation, for an average cluster S/N ratio of 10.3. 
Finally, we study the residuals obtained by extrapolating the track 15~mm upstream onto the 
first telescope layer, which resembles the extrapolation of a particle track back to 
its production point. This has a Gaussian width of (5.7$\pm$0.1)~$\mu$m while simulation 
predicts (5.6$\pm$0.1)~$\mu$m as shown in Figure~\ref{fig:residuals}. 
In order to measure the dependence of these residuals on the S/N ratio of the hit clusters, 
we vary the threshold on the seed pixel S/N ratio of all associated hits and compute the 
r.m.s. width of the residual distribution on the first layer as a function of the average 
S/N ratio of the hits on a track. Results are shown in Figure~\ref{fig:residuals_sn} 
for data and simulation. We also evaluate the effect of hit density on the non-Gaussian 
tails of the residual distribution. The fraction $F$ of tracks with a residual on the 
first layer, which is more than 2.5~$\sigma$ away from zero, is studied as a function of 
the minimum distance, $d_{\mathrm{ch}}$, between the hits associated to the track and the 
closest, not associated hit on any of the layers 2, 3 and 4. The most probable value of 
this distance in data is 0.19~mm. As a comparison, in the core of a $b$ jet from 
$e^+e^- \to Z^0H^0$, $H^0 \to b \bar b$ at 0.5~TeV, disregarding the machine-induced 
pair background, the most probable value of $d_{ch}$ is 0.40~mm. The fraction $F$ is 
0.040, for data where the distance $d_{\mathrm{ch}}$ is larger than 100~$\mu$m, 
and increases to 0.050 and 0.085, when $d_{\mathrm{ch}}$ drops below 100~$\mu$m and 
75~$\mu$m, respectively. This trend is consistent with that predicted by simulation and 
gives a measure of the rate of outliers in the pattern recognition in presence of large 
hit occupancy. 
The extrapolation resolution on the first layer can be extracted from the measured width 
of the residual distribution by subtracting in quadrature the detector resolution, 
taken from simulation. 
This gives an extrapolation resolution of $\sigma_{{\mathrm{extr}}}$=(4.9$\pm$0.1)~$\mu$m 
and (4.2$\pm$0.3)~$\mu$m for cluster S/N values of 10 and 15, respectively. In an earlier 
paper, we reported the measurement of the extrapolation resolution for 1.5~GeV electrons, 
where we obtained $\sigma_{{\mathrm{extr}}}$=(8.5$\pm$0.4)~$\mu$m with the TPPT-1 
telescope~\cite{Battaglia:2006tf}. 
\begin{figure}
\begin{center}
\epsfig{file=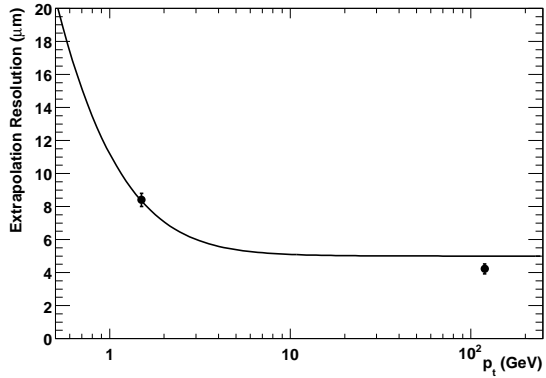,width=7.5cm,clip}
\end{center}
\caption[]{Track extrapolation resolution, $\sigma_{{\mathrm{extr}}}$, as a function of 
the particle momentum. Data at 1.5~GeV and 120~GeV are shown as points with error bars 
and the $5 \oplus \frac{10}{p_t}$~($\mu$m) requirement as a continuous line.}
\label{fig:extrap}
\end{figure}
These results are now compared in Figure~\ref{fig:extrap} to the required track extrapolation 
resolution for $e^+e^-$ collisions at 0.5~TeV at the ILC. These data reaches the 
target performance required for the ILC, using a standalone Si pixel tracker of similar 
geometry to that proposed for an ILC detector, in a high density track environment and 
rather realistic operating conditions with airflow cooling.
 
\section{Vertexing Results}

\begin{figure}
\begin{center}
\begin{tabular}{c}
\epsfig{file=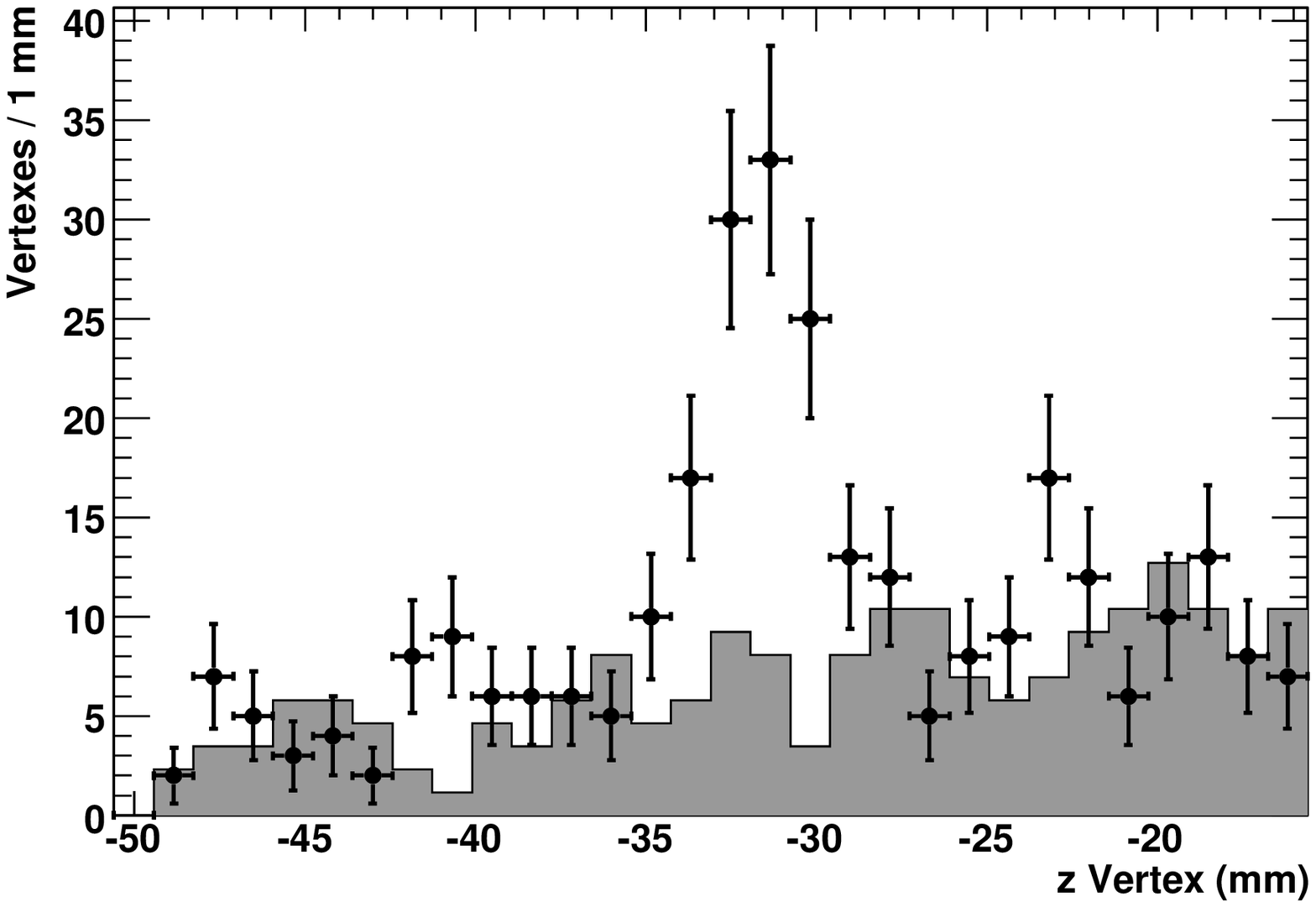,width=7.5cm} \\
\vspace*{-0.5cm} \\
\epsfig{file=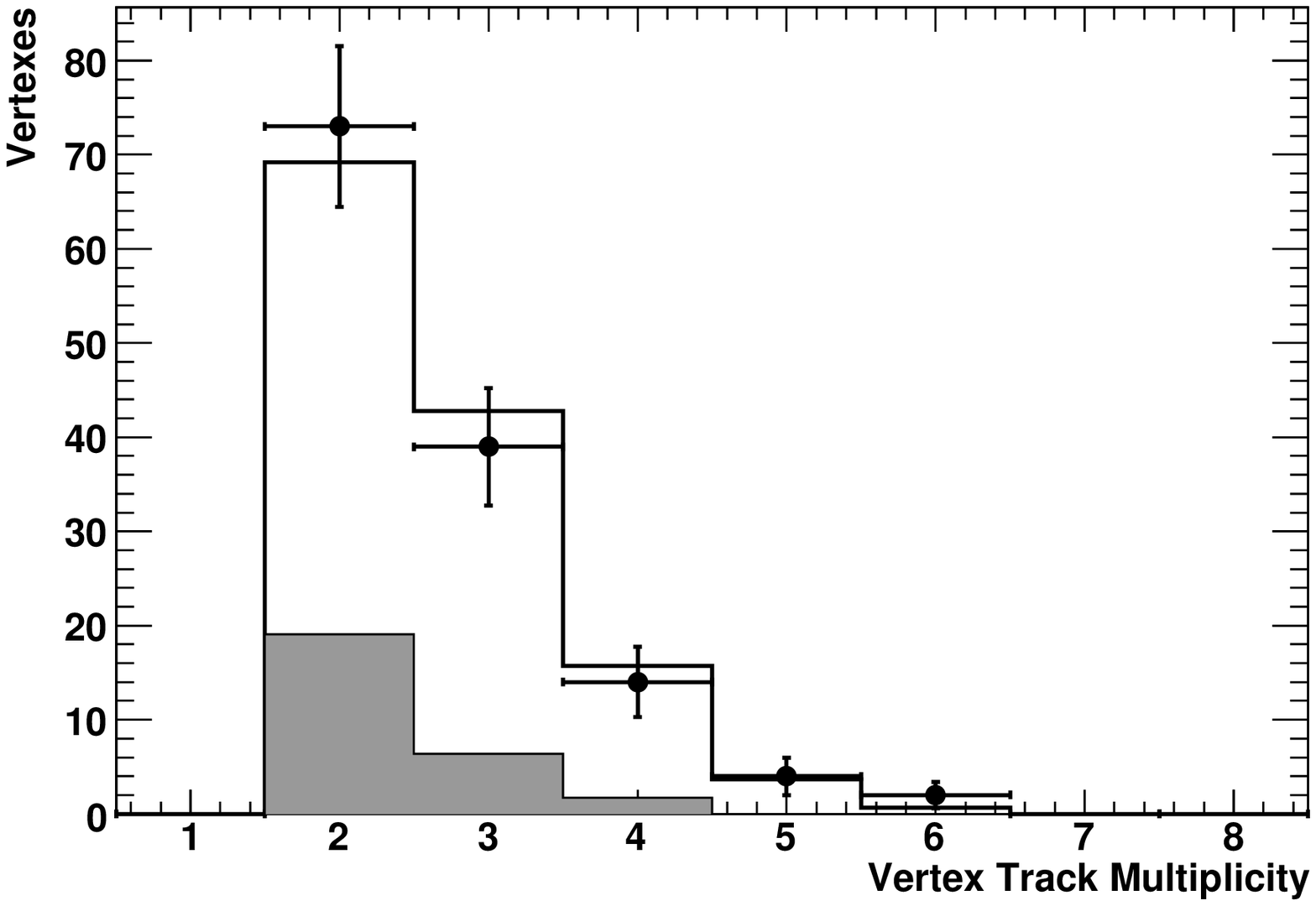,width=7.5cm} \\
\end{tabular}
\end{center}
\caption[]{Vertex reconstruction for p-Cu interaction events: 
(above) reconstructed vertex position along the beam axis, $z$, and (below)
multiplicity of tracks associated to a reconstructed vertex. Data are shown 
as points with error bars and simulation with the open histogram, while the grey filled 
histogram gives the distribution from combinatorial background from events taken 
without target, rescaled to the same number of analysed events as those with the 
target.}
\label{fig:vertex}
\end{figure}

A set of data has been taken with a 4~mm thick Cu target inserted on
the beam-line, $\simeq$~30~mm upstream of the first beam telescope plane. 
These data are used to collect inclined tracks, from protons scattered in 
the target, for alignment purposes as well as for a first study of vertex 
reconstruction performance with a thin pixel tracker telescope in 120~GeV 
$p$-Cu interactions. 
For this study we relax the requirement on the minimum S/N ratio of the seed pixel 
in the cluster to 4 and we raise the maximum track slope to 0.0085. Vertices are 
fitted using a port to the {\tt Marlin} framework of the {\tt VT}~\cite{vtlib1} 
Kalman filter vertex fitter, developed for HERA-B and LHCb, according to the following 
procedure. First a seed vertex is searched for. This is defined as a vertex made of 
a pair of tracks with at least three associated hits, including one on the first layer, 
and having a slope in excess of 0.00075. These cuts suppress combinatorics due to 
either fake tracks or  primary protons not interacting with the target and have 
been optimised using fully simulated and reconstructed p-Cu interaction events. 
Seeds with a fit $\chi^2$ below 20 are kept. 
\begin{figure}
\begin{center}
\epsfig{file=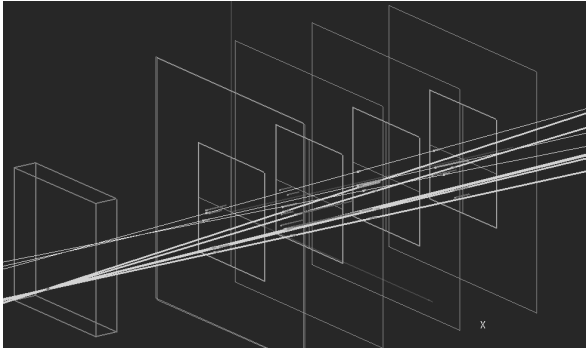,width=7.75cm}
\end{center}
\caption[]{Display of one of the largest multiplicity interaction events:
a reconstructed six-prong vertex from the interaction of an 120~GeV proton 
in the Cu target.}
\label{fig:evtdisplay}
\end{figure}
Additional tracks are tested for association with this seed vertex. 
Tracks are retained if their $\chi^2$ contribution does
not exceed 10 and the vertex position along $z$ does not change by more than 
five times the estimated uncertainty. Finally, unassociated hits are tested for
their compatibility with this vertex. A track search is performed 
using the seed vertex as a track point. Tracks reconstructed with at least two associated 
hits, $\chi^2 <$ 15 and slope below 0.0100 are accepted. The vertex is then refitted 
using all associated tracks. Vertices with $\chi^2/{\mathrm{n.d.o.f.}} <$ 2.5 and at least 
one track with slope in excess of 0.010 are kept. The distribution of the reconstructed 
vertex positions along the beam axis, $z$, shows a clear accumulation at the target location.
An event with a reconstructed six-prong vertex in the target is shown in 
Figure~\ref{fig:evtdisplay}. The analysis has been applied to both simulated events, 
where a comparable number of protons per event collide against a copper target as in the 
actual data, and to data taken without the target in front of the telescope. These data are 
used as a control sample to estimate the background from combinatorics. The data control 
sample has a flat distribution of the $z$ position of the reconstructed vertices, consistent 
with the sidebands of the data target sample (see Figure~\ref{fig:vertex}).
The number of tracks associated to these vertices in data agrees with that predicted by 
simulation for real vertices (see Figure~\ref{fig:vertex}). The average multiplicity of 
tracks at the vertex is 2.74$\pm$0.09 in background-subtracted data and 2.73$\pm$0.04 in 
signal simulation.

Finally, we study the distribution of the resolution on the vertex position along the 
particle line of flight. The estimation of the resolution is validated using simulated 
events. The $z$ pull distribution, i.e.\ the difference between the simulated and 
reconstructed $z$ positions normalised to their estimated uncertainty, has a Gaussian 
width of 1.03$\pm$0.05. The average resolution in 
data is (260$\pm$10)~$\mu$m for vertices $\simeq$32~mm away from the first sensitive 
layer. This can be compared to the resolution obtained for $B$ decay vertices in 
simulated $b$ jets at the ILC, which are $\simeq$15~mm away from the first sensitive layer. 
Simulated vertices from $B$ mesons with energies in the range 100~GeV-150~GeV, 
re-weighted by the associated track multiplicity to match that of the reconstructed 
vertices in our data, have a resolution of (170$\pm$20)~$\mu$m.

\section{Conclusions}

The analysis of the data collected in the first run of the T966 
beam test experiment at FNAL MTBF with a beam telescope based on 
thin monolithic CMOS pixel sensors has provided information on 
tracking and vertexing capabilities in rather realistic conditions.
The resolution for extrapolating tracks 15~mm upstream of the first 
telescope plane is measured to be 4.2~$\mu$m to 4.9~$\mu$m, depending 
on the S/N of the hit clusters used for reconstructing the tracks. 
Vertices from 120~GeV p-Cu interactions have been reconstructed with 
a resolution of (260$\pm$10)~$\mu$m along the flight direction.

\section*{Acknowledgements}

\vspace*{-0.1cm}

This work was supported by the Director, Office of Science, of 
the U.S. Department of Energy under Contract No.DE-AC02-05CH11231.
We are indebted to E.~Ramberg and the Main Injector staff for
their assistance and the excellent performance of the machine 
and to FNAL for hospitality. We gratefully acknowledge the 
contribution of I.~Childres, K.~Deck and M.~Marins De Castro Souza 
to the data taking and data quality checking.

\vspace*{-0.1cm}

 \nolinenumbers


\begin{thebibliography}{99}

\bibitem{Djouadi:2007ik}
  A.~Djouadi, J.~Lykken, K.~Monig, Y.~Okada, M.~J.~Oreglia and S.~Yamashita (Editors),
  {\it International Linear Collider Reference Design Report}, Volume 2: 
  {\it PHYSICS AT THE ILC}, arXiv:0709.1893 [hep-ph].

\bibitem{Battaglia:2004mw}
  M.~Battaglia, A.~De Roeck, J.~R.~Ellis and D.Schulte (Editors),
 {\it Physics at the CLIC multi-TeV linear collider}, CERN-2004-005
 [arXiv:hep-ph/0412251].

\bibitem{vxd3}
  N.~Sinev {\it et al.},
  Nucl.\ Instrum.\ Meth.\  A {\bf 409} (1998) 243.

\bibitem{Battaglia:2006tf}
  M.~Battaglia, D.~Contarato, P.~Giubilato, L.~Greiner, L.~Glesener and B.~Hooberman,
  Nucl.\ Instrum.\ Meth.\  A {\bf 579} (2007) 675.

\bibitem{ilcvtx}
 A.~Besson {\it et al.}, Nucl.\ Instrum.\ and Meth.\ A {\bf 568} (2006) 233.

\bibitem{mimosa}
 Yu.~Gornushkin {\it et al.}, Nucl.\ Instrum.\ and Meth.\ A {\bf 513} (2003) 291.

\bibitem{deptuch-ref}
 G.~Deptuch, Nucl.\ Instrum.\ and Meth.\ A {\bf 543} (2005) 537.

\bibitem{aptek-ref}
 Aptek Industries, San Jose, CA 95111, USA.

\bibitem{Gaede:2003ip}
  F.~Gaede, T.~Behnke, N.~Graf and T.~Johnson,
in the {\it Proc. of 2003 Conf. for Computing in High-Energy and Nuclear Physics}
(CHEP 03), La Jolla, California, 24-28 Mar 2003, pp TUKT001,
  [arXiv:physics/0306114].

\bibitem{Battaglia:2003kn}
  M.~Battaglia,
  Nucl.\ Instrum.\ Meth.\  A {\bf 530} 33 (2004).

\bibitem{Buesser:2005iu}
  K.~Buesser,
{\it In the Proceedings of 2005 International Linear Collider Workshop (LCWS 2005), Stanford, California, 18-22 Mar 2005, pp 1116}.

\bibitem{Gaede:2006pj}
  F.~Gaede,
  Nucl.\ Instrum.\ Meth.\ A {\bf 559} (2006) 177.

\bibitem{Agostinelli:2002hh}
  S.~Agostinelli {\it et al.},
  Nucl.\ Instrum.\ Meth.\ A {\bf 506} (2003) 250.

\bibitem{Battaglia:2007eu}
  M.~Battaglia,
  Nucl.\ Instrum.\ Meth.\  A {\bf 572} (2007) 274.

\bibitem{vtlib1}
  T.~Lohse, Report DESY~95-103.

\end{thebibliography}
\end{document}